# Title: Optical Selection Rule based on Valley-Exciton Locking for 2D Valleytronics

Jun Xiao,[1] Ziliang Ye,[1] Ying Wang,[1] Hanyu Zhu,[1] Yuan Wang,[1,2] Xiang Zhang[1,2,*]

[1] NSF Nano-scale Science and Engineering Center (NSEC), 3112 Etcheverry Hall, University of California at Berkeley, Berkeley, California 94720, USA.

[2] Material Sciences Division, Lawrence Berkeley National Laboratory, 1 Cyclotron Road, Berkeley, California 94720, USA.

**Optical selection rule fundamentally determines the optical transition between energy states in a variety of physical systems from hydrogen atoms to bulk crystals such as GaAs[1,2]. It is important for optoelectronic applications such as lasers, energy-dispersive X-ray spectroscopy and quantum computation[3,4]. Recently, single layer transition metal dichalcogenide (TMDC) exhibits valleys in momentum space with nontrivial Berry curvature and excitons with large binding energy [5-10]. However, it is unclear how the unique valley degree of freedom combined with the strong excitonic effect influences the optical excitation. Here we discover a new set of optical selection rules in monolayer $WS_2$, imposed by valley and exciton angular momentum. We experimentally demonstrated such a principle for second harmonic generation (SHG) and two-photon luminescence (TPL). Moreover, the two-photon induced valley populations yield net circular polarized**

**photoluminescence after a sub-ps interexciton relaxation ($2p \to 1s$) and last for 8 ps. The discovery of this new optical selection rule in valleytronic 2D system not only largely extend information degrees but sets a foundation in control of optical transitions that is crucial to valley optoeletronic device applications such as 2D valley-polarized light emitting diodes (LED), optical switches and coherent control for quantum computing[11,12].**

Optical selection rule is a fundamental principle dictating allowable and forbidden transitions. It is strictly imposed by various symmetries, such as time translational, spatial translational and rotational symmetry, and their corresponding conservations in energy, momentum and angular momentum according to Noether's theorem. In particular, the angular momentum contributed from electron's orbit and spin is important to uncover symmetry of electronic states in atoms using atomic emission spectroscopy, control semiconductor LED and laser polarization, and optical manipulation of spin polarization in spintronics[13,14]. In conventional selection rule studies in solid such as bulk GaAs and its quantum well, angular momentum of Bloch electrons are believed exclusively inherit from their atomic orbits[2]. The role of intercellular electron motion in selection rule received little attention, due to the requirement of inversion symmetry breaking and optical transitions off Brillion zone center [15], until recent discovery of valley angular momentum in monolayer TMDC which has direct bandgap at Brillion zone edge and lacks of inversion symmetry[5-7].

Valley angular momentum (VAM) is associated with energy valleys in momentum space in monolayer TMDC. The inversion symmetry breaking makes it possible that VAM of Bloch electrons contains contributions from not only individual atomic orbits but also the electron circulation from one atom site to another throughout the crystal unit cell. Bloch electrons have opposite signs of VAMs in adjacent valleys and render a valley dependent optical selection rule in linear spectrum[7-9]. Recent experiment reported monolayer WS$_2$ has strong excitonic effect due to reduced dielectric screening[8,16]. Similar to hydrogen atom, the exciton confined in 2D plane also has excitonic angular momentum (EAM) resulting from orbital motion of electron relative to hole. Here we show a new optical selection rule based on valley-exciton locked relationship imposed by EAM and VAM together in monolayer WS$_2$. Our experiments on SHG and TPL confirmed this principle. Such a new selection rule in 2D system is fundamental in determining optical transitions and encoding information not only in valleytronics[17] but also with multiple excitonic degrees.

The three-fold rotation symmetry in monolayer WS$_2$ requires total angular momentum conservation during the light matter interaction, where the VAM, EAM, lattice angular momentum and photon spin angular momentum are exchanged with each other. Induced from the local atom orbital angular momentum and nontrivial Berry curvature distribution, VAM has an out-of-plane component in both conduction ($\tau\hbar$=1 $\hbar$/-1 $\hbar$ at K'/K valleys) and valence ($\tau\hbar$=0 $\hbar$ at K'/K valleys) bands. Meanwhile, the electron-hole relative motion is confined in the 2D plane, resulting in the exciton wavefunction in the form of $R_{n,l}(\rho)e^{il\varphi}$ with only out-of-plane $l\hbar$ EAM, where φ is the azimuthal angle and ρ is the electron-hole distance[18,19]. These out-of-plane VAM and EAM are combined collinearly. In addition, the crystal transforms the impinging angular momentum into a module of three by absorbing the extra angular momentum into lattices, similar to Umklapp process in phonon scattering[20]. As a result, with normal incidence light, the out-of-plane angular momentum conservation yields optical selection rule accounting for the unique VAM and EAM in monolayer WS$_2$:

$$\Delta m\hbar = \Delta\tau\hbar + \Delta l\hbar + 3N\hbar, \quad (N \text{ is integer}) \qquad (1)$$

Upon absorption, spin angular momentum change of photons $\Delta m\hbar$ results in angular momentum change of valley ($\Delta\tau\hbar$), exciton ($\Delta l\hbar$) and crystal lattice ($3N\hbar$). Though strong spin-orbit coupling is present, the spin of electron does not flip under the dominant electrical dipole transition. Therefore, the unchanged out-of-plane spin angular momentum has no contribution to equation (1).

Clearly, both TPL and SHG show a valley-exciton locked selection rule as described by equation (1), that excitonic resonant two-photon process only happens within specific valley under pure σ+ or σ- excitation. For example, in 1$s$ resonant SHG process, the transition from ground state to 1$s$ state at K valley requires -1 $\hbar$ VAM change and 0 $\hbar$ EAM change. From equation (1), it could only take place by absorbing two pure σ+ fundamental photons with 2 $\hbar$ and emitting extra 3 $\hbar$ angular momentum into crystal lattice. The virtual 1$s$ exciton in K valley immediately emit an σ- second harmonic (SH) photon (Figure 1a). In time reversal way, only σ- fundamental photons can trigger transition from ground state to 1$s$ state at K' valley and emit σ+ SH photon in resonant SHG process. Therefore, the SHG emission always has opposite helicity as that of incident light. When it comes to 2$p$ resonant TPL, only the transition from ground state to $2p_+$ state with +1 $\hbar$ EAM at K' valley is induced, under pure σ+ excitation. In this transition, the +1 $\hbar$ VAM and +1 $\hbar$ EAM change can be fulfilled by only absorbing two σ+ fundamental photons without lattice contribution. The 2$p$ exciton from above transition relaxes to 1$s$ state and finally emit an σ+ photon (Figure 1b). Again governed by equation (1), with σ- excitation, only TPL in K valley occur in a time reversal way. In contrast to SHG, 2$p$ resonant TPL always show same helicity as that of incident light. In Table I, we summarize the complete allowed and forbidden TPL and SHG transitions from ground state to excitonic states with a strong optical response under pure circularly polarized light incidence. In the following, we performed polarization resolved SHG and TPL spectral experiment to confirm this new valley-exciton locked selection rule.

WS$_2$ monolayer samples are mechanically exfoliated onto SiO2/Si substrate (Supplementary Figure 1). A typical light emission spectrum is shown under the excitation at 1090 nm (1.14 eV) at 20 K by ultrafast laser (Figure 1 c). One emission peak at 2.28 eV is assigned to SHG. The other two peaks observed at 2.05 and 2.09 eV correspond to the neutral exciton and charged exciton emissions. The charged exciton emission is dominant and selected as our TPL signal. These excitonic emitted photon energies are nearly two times higher than that of the excitation photon, and therefore, they can only originate from the two-photon absorption. This is further confirmed in Figure 1c inset. Both the TPL and the SHG show quadratic power dependence, indicating the two-photon nature.

From equation (1), the SHG carries opposite helicity while TPL displays same helicity as that of incident light. We firstly experimentally examine the optical selection rule for SHG. An excitation energy scan for SHG signal (Figure 2a) shows a resonance at 2.09 eV, indicating an exciton-enhanced second harmonic generation due to the 1$s$ state. The magnitude of SHG signal at 1$s$ state is enhanced almost one order compared with that under non resonant excitation. We then measure SHG with σ+ excitation (Figure 2b) at 1$s$ state. The SHG helicity, defined as $\frac{I(\sigma+)-I(\sigma-)}{I(\sigma+)+I(\sigma-)} \times 100\%$, is -99% at excitation energy of 1.045 eV. The negative sign here means SHG has the opposite circular polarization to that of fundamental light. Determined by selection rule, such high helicity value is preserved because SHG is an instantaneous process and free of any intervalley scattering process[21], The observation of nearly 100% SHG negative helicity and intensity resonance agree well with valley-exciton locked selection rule we proposed.

In contrast to SHG selection rule at 1$s$ state, TPL displays completely opposite valley dependent selection rule at 2$p$ state. In an excitation energy scan (Figure 3a), a dominant resonance is observed at 1.13 eV corresponding to 2$p$ excitonic peak. Compared with 1$s$ excitonic level, 2$p$ exciton has larger linewidth (~80 meV) and asymmetric shape. To exam the selection rule, we measure the TPL spectrum with σ+ light excitation (Figure 3b). The TPL helicity measures 29.6% at such excitation energy (1.13 eV), and has the same sign as that of the incident light. The helicity changes sign when σ- light pumps the monolayer (See Supplementary Figure 2). More interestingly, we observed a TPL resonance in the emission helicity at 2$p$ peak (Figure 3b inset), which confirms our theory that the EAM sets an additional selection rule on the optical transition. The relatively low helicity value here is due to the strong intervalley scattering at such high energy injection[22]. Away from 2$p$ resonance peak, we always observed nonzero TPL with lower energy excitation in Figure 3 (a), which indicates some non-excitonic states with $p$ component below 2$p$ state. This background cannot come from re-absorption of SHG because the emission helicity is positive.

In the following, we further examine the valley-exciton locked selection rule by measuring the time-resolved TPL. After initial two-photon absorption and formation of 2$p$ valley excitons, two subsequent processes would take place: 2$p$ excitons relax to 1$s$ excitons and corresponding 1$s$ excitons' recombination. If the populations in K and K' valleys are unbalanced, intervalley scattering will involve. Here we conducted time-resolved TPL at 2$p$ resonance and the signal is detected by a synchroscan streak camera with an overall time resolution of 2 ps. Under linear polarization excitation, the intervalley scattering has no net contribution because of equivalent population in two valleys. Therefore, the time-resolved TPL trace only includes relaxation and recombination (Figure 4a). The rise edge after excitation mainly contains the relaxation process from 2*p* to 1*s*, while the recombination process is reflected by the decay part. To capture the essence of the dynamics, we use a two-level rate equation (Supplementary material and Supplementary Figure 4). Typically, process no shorter than one tenth of the system time resolution can be extracted by doing convolution fitting with instrument response function (IRF) [23,24]. In this way we can infer that the inter exciton relaxation time is 600±150 fs and recombination time is 5.0±0.2 ps. In spite of the large energy gap between 2*p* and 1*s* excitonic levels, the relaxation completes within 1 ps, which, unlike in quantum dots, indicates no phonon bottleneck impeding hot carrier relaxation between distantly separated energy levels [25]. It is suspected that 1*s* hot exciton with kinetic energy is generated after 2*p* exciton collides with phonon and cascades to 1*s* exciton edge. The rapid recombination observed is likely attributed to nonradiative recombination such as defect trapping or phonon assisted process[26]. This is confirmed by the relative quantum yield measurement, which shows a quantum yield about 1%. We carefully check the nonradiative channel is not dependent on the excitation power (Supplementary Figure 3), which excludes any exciton-exciton annilation mechanism[27].

For σ+ two-photo excitation at 2*p* resonance, the different polarization emission dynamical curves are shown in Figure 4b. The σ+ TPL from K' valley displays higher intensity than that of σ- TPL from K valley, which again confirms the valley-exciton locked selection rule. Compared with σ+ emission, σ- emission always shows slower decay until populations in two valleys are equal. This decay trend difference results from intervalley scattering, which tends to equalize the exciton populations in the two valleys. From convolution fitting with IRF and recombination time, we may estimate the lifetime of intervalley scattering during relaxation and recombination are 3±1 ps and 8.3±0.5 ps (Figure 4b inset). Exchange interaction between electron and hole can induce both spin flip and momentum change via Coulomb potential scatter. Recent calculations show this process can take place in picosecond range and become more efficient if the exciton carries more energy[28], which may account for the rapid depolarization in our observation.

In summary, we discovered the optical selection rules based on valley-exciton locking in monolayer $WS_2$. This important finding reveals EAM and VAM together fundamentally determine optical transitions in monolayer TMDC. This new selection rule sets an important guidance in manipulating exciton and valley degree of freedom in 2D TMDC. For example, the well-defined excitonic levels located in distinguishable valleys could enable quantum computation, lead to THz valley laser and excitonic circuits based on the 2D material[29,30].

# Methods

**Sample preparations:**

Monolayer WS$_2$ samples are mechanically exfoliated onto 275nm SiO$_2$/Si substrate from CVD synthetic crystal flakes (2d Semiconductors Inc.). The exfoliated monolayers are typically 5-10 micrometers in size and characterized by tools such as AFM, Raman and photoluminescence.

**Polarization resolved Second harmonic generation and Two-photon excitation spectroscopy:**

The excitation light is carried out with an optical parametric oscillator (Spectra Physics, Inspire HF 100) pumped by a mode-locked Ti:sapphire oscillator. The laser pulse width is about 200 fs and repetition rate is 80 MHz. The excitation laser is linear polarized by a 900-1300nm polarization beamsplitter. The transmitted p polarization laser light is converted to circularly polarized light via a broadband Fresnel Rhomb quarter wave retarder. The low temperature experiment is operated in a continuous-flow liquid Helium cryostat equipped with a long working distance 50x objective. The emission signal is detected in the back scattering configuration and analyzed by the Fresnel Rhomb quarter wave retarder followed by a visible range polarizer and finally collected by a cooled CCD spectrometer. For time resolved one-photon and two-photon photoluminescence, the signal passing through bandpass filter with a bandwidth of 30 meV is collected by a synchroscan Hamamatsu streak camera with an overall time resolution of 2 ps. The transmissivity of the optical system is carefully calibrated to evaluate the absolute power level at the focusing plane. The emission spectra are normalized to the square of the focused power, as the excitation is limited to the unsaturated regime. The laser pulse width is measured by a home-built autocorrelator at the focus throughout the scanning range.

**Table 1** Optical selection rule based on valley-exciton locking in 2D TMDC Resonant SHG and TPL are only allowed for excitonic states in specific valleys under pure σ+/σ- light excitation. $\Delta\tau\ \hbar$, $\Delta l\ \hbar$ are VAM, and EAM change from ground state to resonant excitonic states. The photon angular momentum change is $2\ \hbar$ /$-2\ \hbar$ with pure σ+/ σ- illumination. The capital letters A/F indicate that the optical transition is "allowed/forbidden" under pure σ+/ σ- illumination. N/A means "not available", indicating the optical transition experiences negligible transition strength based on parity symmetry analysis. The red color indicates σ+ emission while the blue color represents σ- emission for either SHG or TPL.

**Figure legends:**

**Figure 1 | Schematic diagram of optical selection rule based on valley-exciton locking in 2D TMDC. a,** $1s$ exciton-resonant SHG in K' and K valleys. Pumped by σ+ (or σ-) polarization fundamental photon, electron at valence band in K (or K') valley reaches to virtual state. Within the lifetime of virtual state, a second σ+ (or σ-) photon pumps electron from virtual state to $1s$ real state in K (or K') valley. Immediately, a second harmonic photon with σ- (or σ+) polarization emits. **b,** $2p$ exciton-resonant TPL in K' and K valleys. Under σ+ (or σ-) polarization two-photon excitation, the system transitions from the ground state to real $2p_+$ (or $2p_-$) state with $+1\ \hbar$ (or $-1\ \hbar$) EAM in K' (or K) valley through intermediate virtual state. The $2p$ exciton relaxes to $1s$ excitonic state and emit an σ+ (or σ-) photon. Conduction band continuum and valence band continuum are labeled as CBC and VBC. Solid lines are real excitonic states while dashed lines are virtual states. The red color indicates σ+ polarization and blue color for σ- polarization. **c,** Typical monolayer $WS_2$ emission spectrum pumped by 1.14 eV laser pulse at 20 K. The peaks at 2.09 eV and 2.05 eV are the $1s$ state A neutral exciton and its charged exciton (trion) induced by two photo luminescence (TPL). Inversion symmetry breaking in monolayer

WS2 results in second harmonic generation (SHG) signal at 2.28 eV. In the inset, the power dependences of TPL and SHG are plotted, showing quadratic power dependence.

**Figure 2 | Experimental demonstration of SHG selection rule in monolayer WS2 at 20 K. a,** SHG intensity (black dots) versus SHG excitation energy scan from 0.99 eV to 1.23 eV. A dominant SHG intensity peak is observed at 1.045 eV pump energy which is attributed to 1$s$ resonance. **b,** Polarization resolved SHG for σ+ (red curve) and σ- (black curve) detection by σ+ light pumping. The negative SHG helicity P = -99.0%±0.1% is observed from multiple repeat measurements. The excitation energy is 1.045 eV. The significant negative helicity at 1$s$ resonance confirms rigorous selection rule for SHG.

**Figure 3 | Experimental demonstration of TPL selection rule in monolayer WS2 at 20 K**. **a,** TPL intensity versus TPL excitation energy scan from 1.10 eV to 1.22 eV. A dominant TPL intensity peak resonance is observed under 1.13 eV laser pulse excitation, attributed to 2$p$ resonance. **b,** Polarization resolved TPL for σ+ (red curve) and σ- (black curve) detection under σ+ light excitation. The TPL circular polarization of P =29.6% ± 0.5% is observed based on multiple repeat measurements. The excitation energy is 1.13 eV. The inset plots TPL helicity versus TPL excitation energy scan for the same range with σ+ polarization. A helicity resonance is observed at 2$p$ excitonic level. Besides 2$p$ resonance, TPL helicity decreases as excitation energy increases due to energy-dependent intervalley scattering.

**Figure 4 | Time-resolved valley exciton dynamics in monolayer WS$_2$ at 20 K. a,** Time trace of TPL (red curve) excited by linear polarization laser pulse at 1.13 eV. Compared with instrument response function (IRF) in blue curve, TPL time trace shows observable rise and decay feature. Based on our two-level rate equation model (see supplementary materials), convolution fitting (black curve) indicates inter exciton relaxation time is 600±150 fs and recombination time is 5.0±0.2 ps. **b,** Polarization resolved time trace of TPL excited by σ+ polarization at 1.13 eV. Because of intervalley scattering, σ+ emission (red curve) shows larger intensity and more rapid decay than that of σ- (black curve). Blue and green dashed lines are guide lines for eye. The inset is the convolution fitting (purple curve) for time resolved valley exciton population (σ+ - σ−, dark yellow dot). The intervalley scattering time during relaxation and recombination are estimate to be 3±1 ps and 8.3±0.5 ps. The error bar is obtained from fitting on multiple repeat measurements.

**Table 1:**

|  | $\Delta l$ | $\Delta \tau$ | SHG | TPL |
|---|---|---|---|---|
| $(1s, K)$ | 0 | -1 | A/F | N/A |
| $(1s, K')$ | 0 | +1 | F/A | N/A |
| $(2p_+, K)$ | +1 | -1 | F/F | F/F |
| $(2p_+, K')$ | +1 | +1 | A/F | A/F |
| $(2p_-, K)$ | -1 | -1 | F/A | F/A |
| $(2p_-, K')$ | -1 | +1 | F/F | F/F |
| $(2s, K)$ | 0 | -1 | A/F | N/A |
| $(2s, K')$ | 0 | +1 | F/A | N/A |

**Figure 1**

a 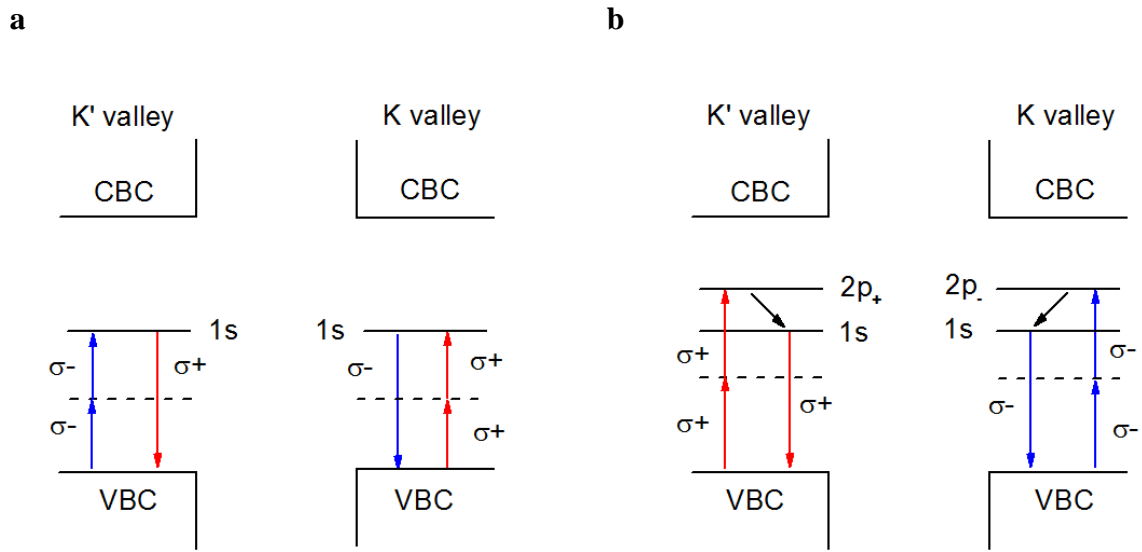 b

c 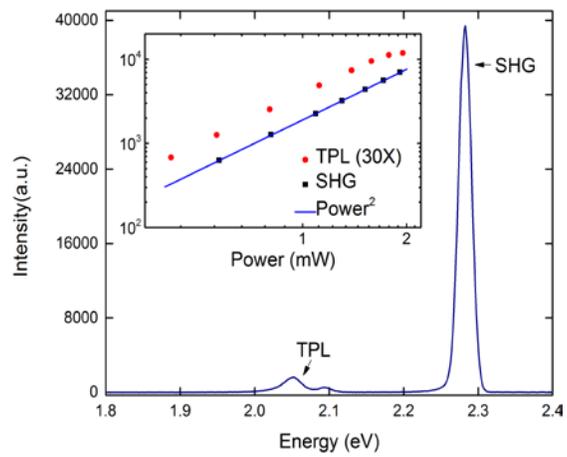

**Figure 2**

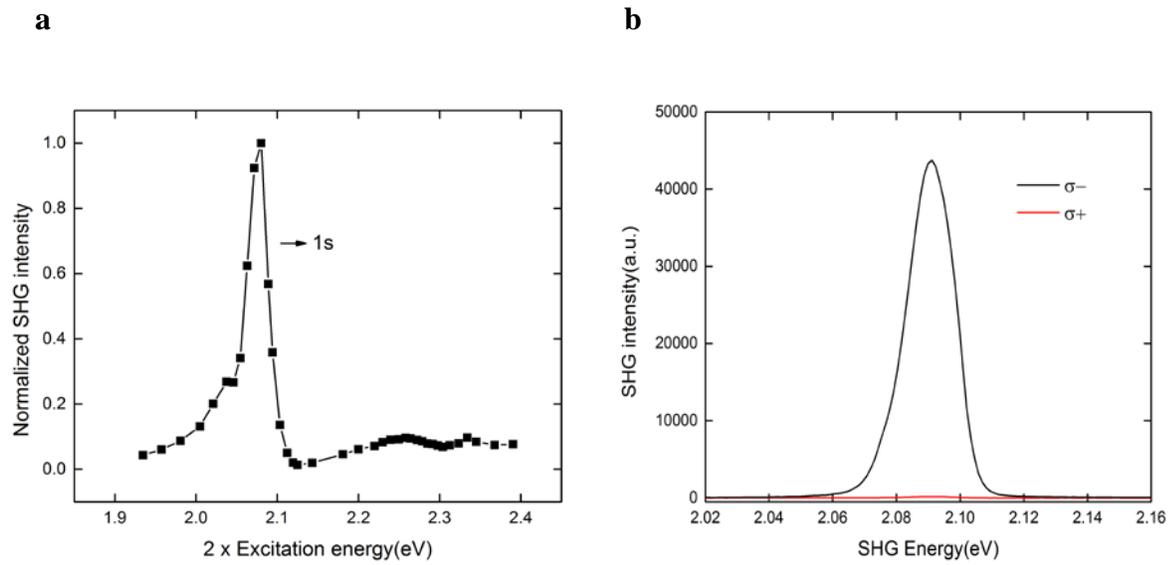

**Figure 3**

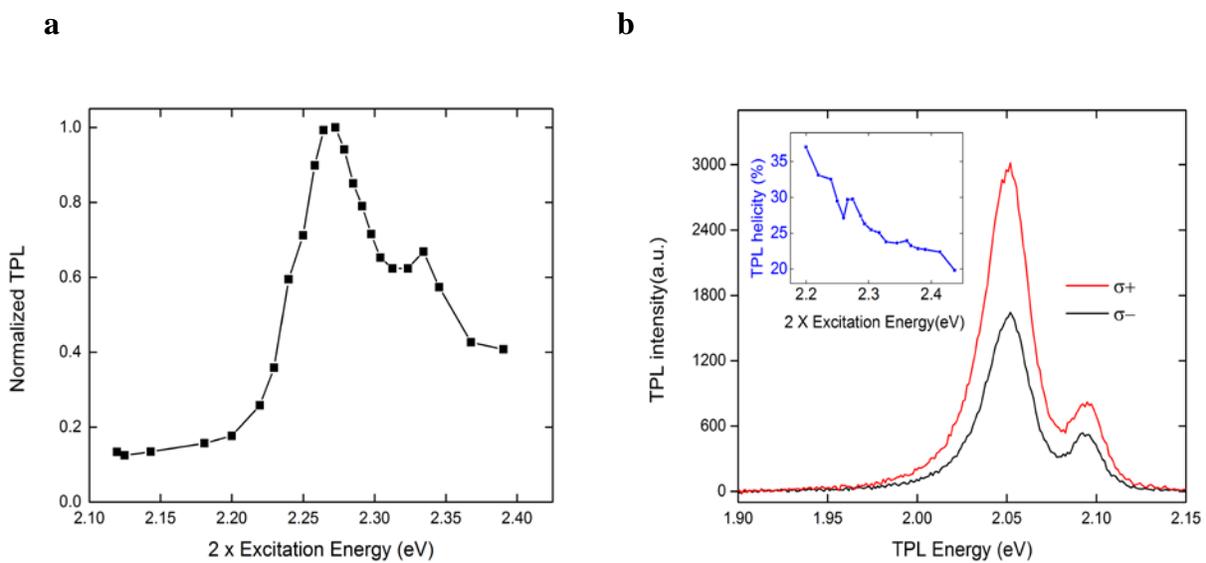

**Figure 4**

a 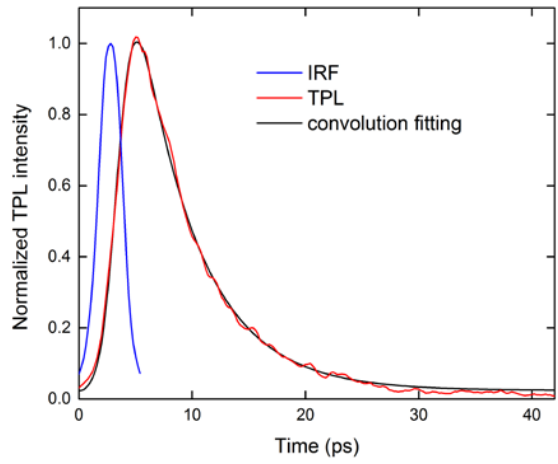 b 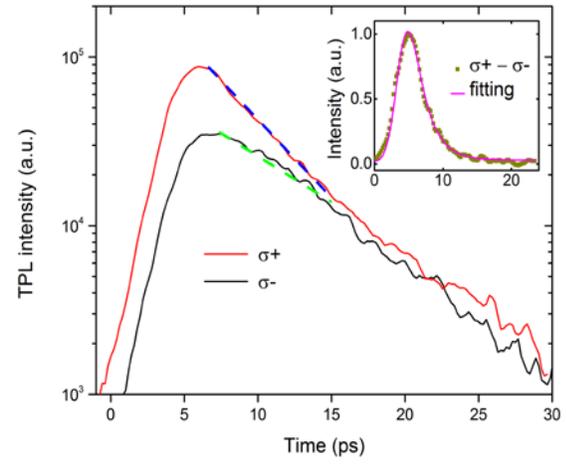